\documentclass[]{article}  
\usepackage{mathptmx}  

\usepackage[T1]{fontenc}
\usepackage{cite}
\usepackage[utf8]{inputenc}
\usepackage{amsmath}
\usepackage{amsfonts}
\usepackage{amssymb}

\usepackage{amsthm}
\newtheorem{theorem}{Theorem}
\theoremstyle{definition}
\newtheorem{property}[theorem]{Property}

\usepackage[cmintegrals]{newtxmath}
\usepackage{tikz}
\usepackage{pgfplots}
\usepackage{multirow}
\usepackage{pgffor}
\usepackage[colorinlistoftodos,prependcaption]{todonotes}
\usepackage{regexpatch}
\usetikzlibrary{shapes,arrows,positioning,patterns,shapes.geometric,lindenmayersystems,shadings}
\usetikzlibrary[shapes.arrows]
\usetikzlibrary{shapes.geometric}
\usetikzlibrary{backgrounds}
\usetikzlibrary{positioning}
\usetikzlibrary{calc}
\usetikzlibrary{intersections}
\usetikzlibrary{fadings}
\usetikzlibrary{decorations.footprints}
\usetikzlibrary{patterns}
\usetikzlibrary{shapes.callouts}
\usetikzlibrary{fit}
\usetikzlibrary{mindmap,trees,positioning,shapes,calc,fadings,arrows,shapes.arrows,shadows,backgrounds, positioning}
\tikzset{lineAlice/.style={thick, black}}
\tikzset{lineBob/.style={thick, black}}
\tikzset{lineMallory/.style={thick, red}}
\tikzset{lineAliceMsg/.style={font=\small,thick, black, draw,->}}
\tikzset{lineBobMsg/.style={font=\small,thick, black, draw,->}}
\tikzset{lineMalloryMsg/.style={font=\small,thick, red, draw,->}}
\tikzset{bubbleAlice/.style={rectangle, draw=black,rounded corners,fill=white,align = flush center,minimum height=0.7cm,minimum width=1.5cm}}
\tikzset{bubbleBob/.style={rectangle, draw=black,rounded corners,fill=white,align = flush center,minimum height=0.7cm,minimum width=1.5cm}}
\tikzset{bubbleMallory/.style={rectangle, draw=red,rounded corners,fill=red,align = flush center,minimum height=0.7cm,minimum width=1.5cm}}
\tikzset{bubble/.style={rectangle, rounded corners,fill=#1,align = flush center,minimum height=0.7cm,minimum width=1.5cm}}
\tikzset{FunctionBubble/.style={draw, fill=blue!30, circle, node distance=2cm}}
\tikzset{DecisionBubble/.style={draw, fill=blue!30, diamond, text badly centered,node distance=2cm}}
\tikzset{ChannelBubble/.style={draw, fill=blue!20, rectangle,minimum height=3em, minimum width=6em}}
\tikzset{bubble/.style={font=\footnotesize,rectangle, rounded corners,fill=white,align = flush center,minimum height=0.7cm,minimum width=1.5cm}}
\tikzfading [name=arrowfading, top color=transparent!0, bottom color=transparent!95]
\tikzset{arrowfill/.style={top color=gray!20, bottom color=gray, general shadow={fill=black, shadow yshift=-0.8ex, path fading=arrowfading}}}
\tikzset{arrowstyle/.style={draw=gray,arrowfill, single arrow,minimum height=#1, single arrow,single arrow head extend=.2cm,}}
\tikzstyle{pinstyle} = [pin edge={to-,thick,black}]
\makeatletter
\xpatchcmd{\@todo}{\setkeys{todonotes}{#1}}{\setkeys{todonotes}{inline,#1}}{}{}
\makeatother

\begin{document}

\title{Impossibility of Three Pass Protocol \\using Public Abelian Groups}

\author{Cansu Betin Onur$^1$ \and Adnan K{\i}l{\i}\c{c}$^2$ \and Ertan Onur$^2$\\
$^1$ Department of Mathematics\\At{\i}l{\i}m University, Ankara Turkey\\cansu.betin@atilim.edu.tr \\
$^2$ Department of Computer Engineering\\Middle East Technical University, Ankara, Turkey\\
\{adnan,eonur\}@ceng.metu.edu.tr}

%\institute{Cansu Betin Onur \at
%              Department of Mathematics, At{\i}l{\i}m University, Ankara Turkey\\
%              \email{cansu.betin@atilim.edu.tr}           %  \\
%%             \emph{Present address:} of F. Author  %  if needed
%           Adnan K{\i}l{\i}\c{c}  \at
%             Department of Computer Engineering, Middle East Technical University, Ankara, Turkey\\
%             \email{adnan@ceng.metu.edu.tr}
%           Ertan Onur  \at
%             Department of Computer Engineering, Middle East Technical University, Ankara, Turkey\\
%             \email{eronur@metu.edu.tr}
%}

\maketitle

\begin{abstract}
Key transport protocols are designed to transfer a secret key from an initiating principal to other entities in a network. The three-pass protocol is a key transport protocol developed by Adi Shamir in 1980 where Alice wants to transport a secret message to Bob over an insecure channel, and they do not have any pre-shared  secret information. In this paper, we prove the impossibility of secret key transportation from a principal to another entity in a network by using the three pass protocol over public Abelian groups. If it were possible to employ public Abelian groups to implement the three-pass protocol, we could use it in post-quantum cryptography for transporting keys providing information theoretic security without relying on any computationally difficult problem. 
\end{abstract}
%\keywords{Key transportation; Three-pass protocol; Cryptography; Group Theoretic Cryptography;  Abelian Groups; Communication Security in Networks} 

\section{Introduction}

Confidentiality in secure communications is defined as ensuring that an
adversary gains no intelligence from a sent message. Alice and Bob would
like to communicate with each other. However, they do not share any secrets.
They only share the endpoints of a communication channel that is fast,
albeit insecure. Anything put on the channel may be tapped by a passive
eavesdropper, Eve. We assume there is no active attacker (e.g., Mallory) in the
system when Alice and Bob talk to each other; i.e., active attacks such as the man  in the middle attack is out of the scope of the paper. Is it possible for Alice and
Bob to exchange messages in a \textit{confidential} manner without having pre-shared secrets (keys)? The answer to this question falls in the scope of key
transport or agreement protocols over insecure channels.

Key transport protocols are designed to transfer a secret key from an
initiating principal to another entity in a network. The initiator
determines the key. However, all of the principals taking part in the
protocol influence the key establishment process in key agreement protocols.
There are many key agreement protocols proposed in the literature that rely
on computationally difficult problems (or in general public-key systems).
The hallmark is the Diffie-Hellman key exchange protocol that makes use of
the difficulty of discrete logarithm problem over a Galois field~\cite{DiffieHellman1976}. It  is widely employed for exchanging secret keys on the Internet~\cite{Xie2016}. Additionally, Ko-Lee-Cheon-Han-Kang-Park~\cite{Ko2000} and Anshel-Anshel-Goldfeld~\cite{anshel1999algebraic} key agreement protocols rely on the computationally difficult conjugacy problem~\cite{myasnikov2011non}.

The three-pass protocol, also known as Shamir's no-key protocol (protocol 12.22 in \cite{menezes1996handbook}), is a key transport protocol developed by Adi Shamir in 1980 where Alice wants to transport a secret message $m$ to Bob over  an insecure channel, and they do not share any secret information. They have mutually agreed on a symmetric encryption scheme that is a pair of
encryption and decryption algorithms $(E,D)$ acting on a message space $\mathcal{M}$ (where $| \mathcal{M}|>1$) and a key space $\mathcal{K}$ such that for all messages $m \in \mathcal{M}$ and keys $k \in \mathcal{K}$, $Prob\left\{ D_k(E_k(m)=m \right\} =1$ where $E_k(m)$ is the notation for encryption of message $m$ with key $k$. Both Alice and Bob determine their secret keys $k_{a}$ and $k_{b}$ respectively without sharing with each other. Alice encrypts $m$ with her secret key $k_a$ and sends $c_1=E_{k_a}(m) $ to Bob. Bob does not have any idea about $k_{a}$, therefore, cannot possibly decrypt $c_1$. Bob sends back to Alice $c_2=E_{k_b}(c_1)$. If $E$ and $D$ commutes, Alice can produce $c_3=D_{k_a}(c_2)=E_{k_b}(m)$ and send it to Bob. Finally, Bob can decrypt it and retrieve the secret message $m$.

If it were possible to employ public Abelian groups to implement the three-pass protocol as we present in Section~\ref{sect:3pass}, we could use it in post-quantum cryptography for transporting keys providing information theoretic security without relying on any computationally difficult problem. As the main contribution of this work, we prove  in Section~\ref{sect:publiccommgroups} that it is impossible to communicate without sharing secrets using three-pass protocol over public Abelian groups. We draw the conclusion in Section~\ref{sect:conclusion} that one has to rely on computational security instead of information-theoretic security to implement the three-pass protocol.  

\section{Three Pass Protocol Using Abelian Groups}
\label{sect:3pass}
Let $(G,\ast )$ be an Abelian group acting on a set $S.$ We apply three-pass
protocol using the action of $G$ on $S$.  If there exists a group
homomorphism 
\begin{align*}
\varphi :G& \rightarrow Sym(S) \\
g& \rightarrow \varphi _{g}
\end{align*}%
where $Sym(S)$ is the group of all permutations on the set $S$ under
composition, then we say that $G$ acts on $S.$ Here, the map $\varphi$  is
called the permutation representation of $G$ on $S.$ To simplify the
notation, we prefer to use the right action notation $a\circ g$ to represent 
$\varphi _{g}(a).$ Note that a group action satisfies $a\circ (g\ast
h)=(a\circ g)\circ h$ and $a\circ 1=a,$ where $1$ is the identity element of 
$G.$ Encryption is defined as the group action. 

We assume that Alice and Bob have agreed on the public Abelian group $G$ and the public homomorphism $\varphi$. Notice that, $G$ and $\varphi$ have to be \textbf{public} since the three-pass protocol assumes no pre-shared information. As shown in Fig.\ref%
{fig:threepassprotocol}, Alice selects the to-be-transported secret key $k \in S$ and a private
group element $g \in G$ uniform randomly, sends Bob $c_1=k \circ g.$ Notice here that this operation is the encryption of the secret message $k$ using the group action $g$ (i.e., $c_1=E_g(k)$). Bob selects his private
group element $h\in G$ uniform randomly and sends back to Alice $c_2=c_1 \circ h$. Then, Alice
computes $c_3= c_2\circ g^{-1}.$ Following the properties of group actions $%
c_3=((k \circ g) \circ h) \circ g^{-1} = k \circ ( g*h* g^{-1}) .$ Since we
chose an Abelian group, $c_3=k \circ (g*g^{-1}*h)=k \circ (1*h)=k \circ h.$  In this protocol, all the messages conveyed over the channel, $c_1, c_2, c_3 \in C=S$ are encrypted. At the end of these three exchange of messages, Bob
is able to secretly compute the key $k$ by $(k \circ h) \circ h^{-1}=k \circ
1 =k.$

\begin{figure}[t]
\centering
\begin{tikzpicture}[auto, node distance=3cm,>=latex']
 \def \n {4}
 \def \height {4cm}
 \def \width {5cm}
 \def \aliceposx {0}
 \def \aliceposy {0}
 \def \bobposx {\width}
 \def \bobposy {0}

% n coordinates for messaging
\foreach \s in {0,...,\n}
{
  \coordinate (a\s) at (\aliceposx, {\height - (\height / \n) * \s});
  \coordinate (b\s) at (\bobposx, {\height - (\height / \n) * \s});
};
 
% Bubbles for Alice and Bob
\node[bubbleAlice] (A) at (a0) {$Alice$};
\draw[lineAlice] ($(A.south)$)--($(A.south)-(0,\height)$);
\node[bubbleBob] (B) at (b0) {$Bob$};
\draw[lineBob] ($(B.south)$)--($(B.south)-(0,\height)$);

\node[bubble,left=0.4cm of a1]{Pick $k \in S$\\ and $g \in G$};
\node[bubble,right=0.4cm of b1]{Pick $h \in G$};
 \draw [lineAliceMsg] (a1) -- node[above]{ $c_1=k \circ g$} (b1);
 \draw [lineBobMsg] (b2) -- node[above]{$c_2=c_1 \circ h$} (a2);
 \draw [lineAliceMsg] (a3) -- node[above]{$c_3= c_2\circ g^{-1}$} (b3);
\node[bubble,right=0.4cm of b3]{Compute\\$k=c_3 \circ h^{-1}$};
%  
%\begin{pgfonlayer}{background}
%\fill[black!10]($(a0)!0.5!(b0)+(0,0.5)$)rectangle($(a0)!0.5!(b0)+(5,-\height)-(0,1)$) ;
%\fill[black!10]($(a0)!0.5!(b0)+(-5,0.5)$)rectangle($(a0)!0.5!(b0)+(0,-\height)-(0,1)$);
%\end{pgfonlayer}
\end{tikzpicture}
\caption{Three-pass protocol using public Abelian group actions.}
\label{fig:threepassprotocol}
\end{figure}
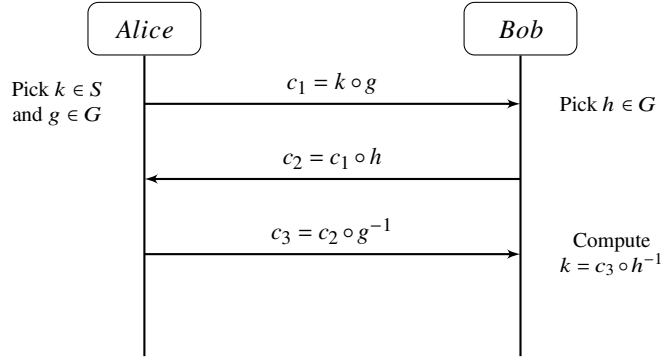

\subsection{Requirements of the Public Group $G$}

$G$ has to satisfy the following properties.

\begin{property}{Commutativity:}
\label{property:comm} 
The group $G$ must be Abelian. If $G$ is not Abelian, then $c_3$ need not to be  equal to $k \circ h.$ 
\end{property}

\begin{property}{Uniform Randomness:}
\label{property:randomness} 
 Let $\mathbf{C}$, $\mathbf{S}$ and $\mathbf{G}
$ be the random variables denoting the values of ciphertexts (e.g., $c_1, c_2
$ or $c_3$ or any $c \in C$), secret keys to be transport (e.g., $k \in S$)
and group elements (e.g., $g, h \in G$), respectively. To be able to provide
information theoretic security, uniform randomness must be satisfied, i.e., 
\begin{equation*}
Prob\left\{\mathbf{C}=c | \mathbf{S}=k, \mathbf{G}=g \right\} = Prob\left\{ 
\mathbf{C}=c | \mathbf{S}=k, \mathbf{G}=h \right\}
\end{equation*}
and 
\begin{equation*}
Prob\left\{\mathbf{C}=c | \mathbf{S}=k_0, \mathbf{G}=g \right\} =
Prob\left\{ \mathbf{C}=c | \mathbf{S}=k_1, \mathbf{G}=g \right\}
\end{equation*}
providing computational indistinguishability for any probability
distribution over the set $S$, the group $G$ and ciphertext space $C=S$
where $Prob\left\{ \mathbf{C}=c \right\}>0.$ Here, we claim that the
probability of conveying any ciphertext over the channel in this protocol is
equally likely. Therefore, eavesdropping does not give any statistical
advantage to an the attacker, i.e., 
\begin{equation*}
Prob\left\{ \mathbf{S}=k | \mathbf{C}=c \right\} = Prob\left\{ \mathbf{S}=k
\right\}.
\end{equation*}
\end{property}
We select the to-be-transported secret key $k \in S$ and a private
group element $g \in G$ shown in Fig.~\ref{fig:threepassprotocol} uniform randomly to satisfy Property~\ref{property:randomness}. Furthermore, for any pair $(a,b)\in S\times S,$ there must
be exactly $M \geqslant 1 $ group elements sending\footnote{When we say $g$ sends $a$ to $b,$ we mean $b = a \circ g.$ } $a$ to $b$. In particular, this action must be transitive as defined in Appendix~\ref{Append:Transitive}. 

\subsection{An Easy Example Implementation}

%There are various Abelian groups. The \textit{easiest} example would be to use $Z_{2}$ as $G$. In this case, the three pass protocol implementation reduces to one-time pad where multiple encryptions with the same key takes place indicating insecurity.

An easy example is to use the natural action of Klein four-group $V$. Klein
four-group is a well-known Abelian permutation group on $S=\{1,2,3,4\}$
whose non-identity elements has order two. That is, every element is its own inverse. It
is a subgroup of $Sym(S).$ For the sake of simplicity and clarity, we
present $V$ using the two-row notation where images are given in the second
row of $\tau =%
\begin{pmatrix}
1 & 2 & 3 & 4 \\ 
\tau (1) & \tau (2) & \tau (3) & \tau (4)%
\end{pmatrix}%
.$ The Klein four-group is $V=\{\tau _{0},\tau _{1,}\tau _{2},\tau _{3}\}$
where $\tau_{0}= 
\begin{pmatrix}
1 & 2 & 3 & 4 \\ 
1 & 2 & 3 & 4%
\end{pmatrix}%
, $
$\tau_{1}= 
\begin{pmatrix}
1 & 2 & 3 & 4 \\ 
2 & 1 & 4 & 3%
\end{pmatrix}%
, $ $\tau_{2}= 
\begin{pmatrix}
1 & 2 & 3 & 4 \\ 
3 & 4 & 1 & 2%
\end{pmatrix}%
, $ and $\tau_{3}=%
\begin{pmatrix}
1 & 2 & 3 & 4 \\ 
4 & 3 & 2 & 1%
\end{pmatrix}%
. $

The implementation of the three-pass protocol using $V$ is as follows. Alice
and Bob will employ $V.$ Suppose Alice selects $k=3$ as the secret key and
her private group element $g=\tau_{1}$ uniform randomly. She sends Bob $%
c_{1}=3\circ \tau_{1} = 4.$ Suppose Bob selects his private group element $%
h=\tau_{2}$ uniform randomly as well. Then, he sends back $c_{2}= 4 \circ
\tau_{2} = 2.$ Alice computes $c_{3}=2 \circ \tau_{1}^{-1}= 2 \circ
\tau_{1}=1$ and sends it to Bob. Finally, Bob recovers the secret key by $%
k=1 \circ \tau_{2}^{-1}=1 \circ \tau_{2} = 3$, which is the secret key determined by Alice following some probability distribution.

$V$ satisfies both Property~\ref{property:comm} and Property~\ref{property:randomness}.   Any message $c_{1},c_{2}$ or $c_{3}$ put on the insecure channel is
encrypted. Therefore, the reader may initially think it is not possible for
Eve to recover the key $k.$ However, this implementation is prone to
brute-force attacks. Therefore an additional requirement is imposed on the
order of $G$:

\begin{property}{Large Order:}
\label{property:order} 
Trivially, the order of the group $G$ and the
cardinality of $S$ must be very large to make the brute force attack almost
infeasible. A brute-force attack can be (theoretically) used against any
cryptosystem to find out the secrets. In this protocol, the private key is
an element of the selected group $G$. Hence, the probability that an
attacker successfully finds the correct element of the group $G$ is equal to 
$1/|G|$.
\end{property}

We present in the sequel that a much deeper security flaw resides in such implementations of the three-pass protocol even if we can find a $G$ that satisfies all these three properties.

%\begin{table*}[tbp]
%\centering
%\begin{tabular}{|p{3cm}|p{3cm}|p{3cm}|p{3cm}|}
%\hline
%Requirements & $Z_2$ & $V$ & Transitive Abelian permutation group \\ \hline
%Abelian &  &  &  \\ \hline
%Uniform randomness &  &  &  \\ \hline
%Order &  &  &  \\ \hline
%Implementation &  &  &  \\ \hline
%Security flaw &  &  &  \\ \hline
%\end{tabular}%
%\end{table*}

\section{Public Commutative Groups and Impossibility}
\label{sect:publiccommgroups}

When Eve observes $c_1=4$ and $c_2=2$ in the above example, she immediately
finds out Bob's private group element $h=\tau_2,$ since there is only one
element $\tau $ in $V$ sending $a$ to $b$ for any pair of elements $(a,b)\in
S\times S.$ It is computationally easy to determine  $\tau $ in $V$ when $a$ and $b$ are given as in the three-pass protocol. In fact, this security flaw is valid for any publicly known transitive
Abelian group $G$ even when the order of $G$ is very large and even when there exist many distinct group elements sending $ a $ to $ b $. Determining one group element sending $ a $ to $ b $ is enough to break the system and determine the secret group elements $g, h \in G$ and the secret key $k$ to be transported shown in Fig. ~\ref{fig:threepassprotocol}. 

One may get this result as follows. Let $G$ be an Abelian group acting transitively on $S$ via a homomorphism $\varphi
:G \rightarrow Sym(S).$ Take any $ a,b $ in $ S. $ Suppose that there exist two distinct group elements $ g_1, g_2 $ sending $ a $ to $ b. $ Then $g_1*g_2^{-1}$ is in the point stabilizer of $a$ (see Appendix~\ref{Append:Stabilizer}). To put it plainly, let $a \circ g_1 = b = a \circ g_2$. Then $a\circ (g_{1}\ast g_{2}^{-1})=(a\circ
g_{1})\circ g_{2}^{-1}=b\circ g_{2}^{-1}=a.$ Therefore, $g_{1}\ast g_{2}^{-1}
$ is an element of $G_{a}$. Say $g_1= \alpha * g_2 $ for some $\alpha \in G_a.  $ Now consider the kernel (see Appendix~\ref{Append:Kernel}) of the permutation representation $\varphi$, which is $Ker \varphi= \bigcap_{x \in S} G_x.$ Take any element $x \in S.$ 
Since, $G$ is transitive on $S$, there exists an element $r \in G$ such that $x=a \circ r$. 
Then by Theorem 1.4A-ii in \cite{DixonMortimer} presented in Appendix~\ref{Append:theorem}, $ G_x=r^{-1} \ast G_a \ast r. $ As $ G $ is Abelian, we get 
$G_x= r^{-1} \ast G_a \ast r=r^{-1} \ast r \ast G_a=G_a$ for all $x $ and so $Ker \varphi= G_a.$ Hence $ \varphi(g_1)= \varphi(\alpha * g_2)=\varphi(g_2).$ For the convenience of the reader, let us verify the last equality elementwise;
 $x \circ g_1=(a\circ r) \circ (\alpha * g_2) = a \circ (r * \alpha * g_2)=
a\circ (\alpha * r *  g_2) = (a \circ r) \circ g_2=x\circ g_2$ since $a\circ \alpha=a.$
 That is for any pair $a,b$ in $S,$ there may be two distinct group elements sending 
$ a $ to $ b $ but in fact the corresponding permutation is unique.

Let's relax the assumptions and consider that the above action of the group $ G $ on the set $ S $ is not transitive. In this case, we cannot satisfy Property~\ref{property:randomness}  since  $Prob\left\{ \mathbf{C}=c \right\}= 0$ for some ciphertext $c$.  Even if we do not get the equality  $ \varphi(g_1)= \varphi(g_2)$ for the elements satisfying $a \circ g_1 = b = a \circ g_2$ the security flaw still exists.
 
 Note that the equality $g_1= \alpha * g_2 $ for some $\alpha \in G_a $ holds whenever $a \circ g_1 = b = a \circ g_2$. 
  As a consequence of this equality, it will be sufficient for the attacker to
find any $g$ that satisfies $a\circ g=b$ instead of trying to find the
selected private group element. In the three pass-protocol, Alice selects a
secret key $k\in S$ and a private group element $g\in G$, sends Bob $%
c_{1}=k\circ g.$ Then Bob selects his private group element $h$ and sends
back to Alice $c_{2}=c_{1}\circ h$. Suppose Eve finds a group element $%
h^{\prime }$\ satisfying $c_{1}\circ h^{\prime }=c_{2}.$ Then $h^{\prime
}=\alpha \ast h$ for some $\alpha \in G_{c_{1}}.$ When Alice sends $%
c_{3}=(c_{1}\circ h)\circ g^{-1}$, Eve is able to recover the secret key by
computing $c_{3}\circ (h^{\prime })^{-1}.$ Indeed, $c_{3}\circ (h^{\prime
})^{-1}=((c_{1}\circ h)\circ g^{-1})\circ (h^{\prime })^{-1}=c_{1}\circ
(h\ast g^{-1}\ast (h^{\prime })^{-1})=c_{1}\circ (h\ast g^{-1}\ast (\alpha
\ast h)^{-1})=c_{1}\circ (h\ast g^{-1}\ast h^{-1}\ast \alpha
^{-1})=c_{1}\circ (h\ast g^{-1}\ast h^{-1}\ast \alpha ^{-1}).$ Note that $%
\alpha ^{-1}\in G_{c_{1}}$ as $\alpha \in G_{c_{1}}.$ Since the group is
Abelian, one has $c_{3}\circ (h^{\prime })^{-1}=c_{1}\circ (\alpha ^{-1}\ast
g^{-1}\ast h\ast h^{-1})=(c_{1}\circ \alpha ^{-1})\circ g^{-1}=c_{1}\circ
g^{-1}=k.$ Therefore, the three-pass protocol is  insecure when public Abelian groups are employed.

\section{Conclusion}
\label{sect:conclusion}
In this paper, we prove  the three-pass protocol is insecure when public Abelian groups are employed. We consider using  Abelian groups since the principals' operations have to be commutative for the three-pass protocol to work. The group has to also be public  since the principals should not have any secret prior to communication. When we assume they have some shared secret information (e.g., an agreement on a secret Abelian group), the innovative feature of the three-pass protocol -being able to establish a secure channel without sharing any secret information in advance - becomes nonsensical. Under these assumptions, it is impossible to securely transport a shared secret by employing the three-pass protocol over public Abelian groups.  One has to rely on computational security instead of information-theoretic security to implement the three-pass protocol as we encounter in many practical key  exchange protocols such as the Diffie-Hellman key exchange protocol that relies on the discrete logarithm problem.

\section{Appendices}

\subsection{Kernel of Group Homomorphism}
\label{Append:Kernel}
Let  $\varphi :G \rightarrow H$  be a group homomorphism. Then the set $ Ker ( \varphi)=\{ x \in G  : \varphi(x)=1_H   \} $ is called the  { \bf kernel } of $ \varphi. $

\subsection{Stabilizer of a Point}
\label{Append:Stabilizer}
Let  $ G $ be a group  acting on a set $ S $ and let  $ a $ be an element of  $ S .$ The set of elements in $ G $ which fix $ a $
 is called the { \bf stabilizer  }  of $ a $ in $ G $   and it is denoted by $ G_a. $

\subsection{Transitive Action}
\label{Append:Transitive}
A group $ G $  acting on a set $ S $ is said to be  { \bf transitive } if for any pair of elements $ a,b \in  S, $ there exists $ x \in G $
 such that $ a \circ x=b. $

\subsection{Theorem 1.4A-ii in \cite{DixonMortimer} }
\label{Append:theorem}
Suppose that $(G,\ast)$ is a group acting on a set $S$ and that $r \in G$ and 
$a, x \in  S.$ Then, the stabilizer $G_a$ is a subgroup of $G$ and $G_x=r^{-1} \ast G_a \ast r$ whenever $x=a \circ r$.

%\bibliographystyle{imaiai}
%\bibliographystyle{plain}
%\bibliographystyle{spmpsci} 
%\bibliography{ref}

\end{document}